\documentclass{article}
\usepackage{spconf,amsmath,graphicx,hyperref,subcaption}
\usepackage{amsfonts, algorithm, algpseudocode, multirow, booktabs}
\usepackage{cite}
\usepackage{rotating}
\DeclareMathOperator*{\argmax}{arg\,max}
\usepackage{color, xcolor}
\usepackage{amsmath,amssymb}

\definecolor{blind_color}{HTML}{00008B}
\definecolor{ours_color}{HTML}{008000}
\definecolor{oracle_color}{HTML}{B22222}

\title{Two-Dimensional Tomographic Reconstruction from Projections with Unknown Angles and Unknown Spatial Shifts}

\name{Shreyas Jayant Grampurohit\textsuperscript{1}, Satish Mulleti\textsuperscript{1}, and Ajit Rajwade\textsuperscript{2}}

\address{\textsuperscript{1}Department of Electrical Engineering, \textsuperscript{2}Department of Computer Science and Engineering \\
         Indian Institute of Technology Bombay, Mumbai, India \\
         \texttt{\{shreyasgrampurohit, mulleti.satish\}@gmail.com, ajitvr@cse.iitb.ac.in}}
\begin{document}
%
\maketitle
\begin{abstract}
In parallel beam computed tomography (CT), an object is reconstructed from a series of projections taken at different angles. However, in some industrial and biomedical imaging applications, the projection geometry is unknown, completely or partially. In this paper, we present a technique for two-dimensional (2D) tomography in which both viewing angles and spatial shifts associated with the projections are unknown. There exists literature on 2D unknown view tomography (UVT), but most existing 2D UVT algorithms assume that the projections are centered; that is, there are no spatial shifts in the projections. To tackle these geometric ambiguities, we first modify an existing graph Laplacian-based algorithm for 2D UVT to incorporate spatial shifts, and then use it as the initialization for the proposed three-way alternating minimization algorithm that jointly estimates the 2D structure, its projection angles, and the corresponding shifts. We evaluate our method on noisy projections of ribosome images and demonstrate that it achieves superior reconstruction compared to the baseline that neglects shifts.
\end{abstract}
\section{Introduction}
\label{sec:intro}

Tomographic imaging is a technique used to determine the internal structure of an object by acquiring Radon projections from multiple angles. It is widely used in various fields, including medicine, materials science, and structural biology \cite{kak2001principles}. In computed tomography (CT), the projection angles are known in advance. However, in some real-world applications, the projection geometry is often not known \cite{basu2000feasibility}, \cite{Goncharov1988-wx}. For example, in applications such as cryo-electron microscopy (CryoEM), the three-dimensional (3D) structures need to be reconstructed from a micrograph containing projections obtained at unknown viewing angles \cite{singer2011cryo}. Moreover, errors in particle-picking (the process of locating the projections on the micrograph) often introduce spatial shifts in the extracted projections \cite{Frank2016-tz}. If these shifts are left uncorrected, they severely degrade reconstruction quality. These challenges are not limited to CryoEM. Accurate angle estimation and shift correction are important for enhancing spatial resolution in high-resolution computed tomography (HRCT) \cite{Lynch2015-gj}. Furthermore, the challenge of unknown viewing angles is also present in the phantomless calibration of CT imaging systems \cite{meng2012online}. As a result, these scenarios demand reconstruction methods that can simultaneously infer the translational shifts, the unknown angles, and recover the underlying image.

In the two-dimensional (2D) unknown-view tomography (UVT) literature, several approaches have been proposed. One common strategy is to first estimate the circular ordering of the noisy projection angles using nonlinear dimensionality reduction algorithms like Laplacian eigenmaps \cite{coifman2008graph} or spherical multidimensional scaling \cite{fang2010estimating}, \cite{fang2011slle}, followed by angle assignment using order statistics of the distribution of the angles (which is assumed to be known). Other approaches include moment-based methods \cite{phan2017moment}, \cite{malhotra2016tomographic}, adversarial learning-based techniques \cite{adversarial2022}, and methods based on signal reconstruction from samples at unknown locations \cite{shah2025signal}. While some recent work has addressed challenges like unknown non-uniform angle distributions \cite{shah2025unknown}, a critical limitation across all these advanced techniques is the assumption that the projections are correctly centered; the issue of unknown spatial shifts is not handled. The dual ambiguities of angle and shift are addressed in \cite{basu2000feasibility}, \cite{basu2000uniqueness} using moment-based methods. The recent work in \cite{Hjouj2025-mr} first centers each projection using its first moment and then compares second- and third-order projection moments to estimate the projection angles. However, such geometric moments are not robust to noise.

In this paper, we propose a robust, two-stage framework to address the challenging problem of tomographic reconstruction from projections with both unknown angles and unknown spatial shifts. The core of our method is a three-way alternating minimization algorithm that jointly refines the underlying 2D image, the projection angles, and their corresponding shifts. Recognizing that the success of such iterative optimization critically depends on a strong initial estimate, we propose a shift-aware initialization scheme. We develop this by modifying a graph Laplacian-based algorithm to make it resilient to the translational ambiguities in the projection data. By combining our robust initialization with the subsequent alternating minimization, our method achieves superior reconstruction quality on noisy, shift-corrupted data compared to a baseline that neglects these geometric uncertainties. While our work assumes a uniform distribution for angle initialization, the framework is flexible and could be extended to incorporate methods for simultaneously inferring the unknown angle distributions, such as in \cite{shah2025unknown}, as a direction for future work.


\section{Problem Formulation}
\label{sec:problem}
Let $f(s, t)$ denote a two-dimensional image defined on $\mathbb{R}^2$. In tomographic imaging, projections of this image are acquired along various directions. The Radon projection of $f$ along the line $s\cos{\theta} + t\sin{\theta} = \rho$ is given by:
\begin{align}
    \mathcal{R}_{\theta}f(\rho) = \iint_{\mathbb{R}^2}f(s, t)\delta(s\cos{\theta} + t\sin{\theta} - \rho) \, ds \, dt,
\end{align}
where $\delta(.)$ is the Dirac delta function, $\theta \in [0, 2\pi)$ is the projection angle and $\rho$ is the perpendicular distance of the projection line from the origin. In many real-world applications, the image undergo an unknown 2D spatial shift $(s_0^i, t_0^i)$ before the projections are taken. The shift theorem of the Radon transform \cite{jain1989} states that this spatial shift in the image results in a corresponding one-dimensional shift in its projection, as expressed below:
\begin{align}
    y_{\theta_i}(\rho) = \mathcal{R}_{\theta_i}f(\rho-\alpha_i).
    \label{subs_eqn_y}
\end{align}
Here $y_{\theta_i}(\rho)$ is the acquired projection and the shift in the projection domain, $\alpha_i$, is given by $\alpha_i := s_0^i\cos{\theta_i}+t_0^i\sin{\theta_i}$.

The objective is to recover the image $f(s, t)$ from a set of its projections $\{y_{\theta_i}(\rho)\}_{i=1}^N$, given that both the projection angles $\{\theta_i\}_{i=1}^N$ and the spatial shifts $\{(s_0^i, t_0^i)\}_{i=1}^N$ are unknown. However, with these unknown parameters, reconstruction is only possible up to a global rotation, reflection, and translation \cite{basu2000uniqueness}. This ambiguity is acceptable for most practical purposes as the exact orientation of the object is not required. We consider a discrete image representation $\boldsymbol{f}$ of size $S \times S$, and denote the set of $N$ acquired projections as 1D vectors $\{\boldsymbol{y_i}\}_{i=1}^N$, where each vector contains $S$ samples. Given $\{\boldsymbol{y_i}\}_{i=1}^N$, the aim is to recover $\boldsymbol{f}$ as well as the unknown angles $\{\theta_i\}_{i=1}^N$ and projection shifts $\{\alpha_i\}_{i=1}^N$ -- upto a global rotation, reflection, and translation. 

\section{Proposed approach for image reconstruction under unknown shifts and angles} 
\label{sec:method}

Our proposed method jointly estimates the image, its projection angles, and the corresponding spatial shifts. The core of our approach is a three-way alternating minimization scheme that iteratively refines these three components. However, the performance of this non-convex optimization is highly dependent on a good initial guess for the angles. Therefore, we also propose a shift-aware initialization technique to provide a robust starting point for the minimization process.

\subsection{Three-Way Alternating Minimization}\label{subsec:twam}
\noindent\textbf{Shift Estimation:} The shifts $\{\hat{\alpha}_i\}_{i=1}^N$ are initialized to zero. In each iteration, we first compute projections, denoted $\{\boldsymbol{\hat{y}}_i := \mathrm{Proj}(\boldsymbol{\hat{f}}, \hat{\theta}_i)\}_{i=1}^N$, of the current image estimate $\boldsymbol{\hat{f}}$ at the current angle estimates $\{\hat{\theta}_i\}_{i=1}^N$ (the initialization of the angles is explained in Sec.~\ref{subsec:initangles}). The shift for the $i$-th projection is then updated by finding the integer shift $k$ that maximizes the dot product between the observed projection $\boldsymbol{y}_i$ and the corresponding re-projection $\boldsymbol{\hat{y}}_i$:
\begin{align}
\hat{\alpha}_i = \argmax_{k} \boldsymbol{y}_i \circledcirc \text{shift}_k(\boldsymbol{\hat{y}}_i),
\end{align}
where $\text{shift}_k(\cdot)$ is an operator that shifts a vector by $k$ indices and $\circledcirc$ denotes the dot product.

\noindent\textbf{Image Reconstruction:} In the image reconstruction step, we first apply the \emph{reverse} of the current projection shift estimates $\{\hat{\alpha}_i\}$ to the given projections to compensate for the spatial shifts. Then, we reconstruct the image using the shifted projections and the current estimate of angles by the well known Filtered Back Projection (FBP) method \cite{Ramachandran1971}.

\noindent\textbf{Angle Updates:} In the angle estimation step, we first apply the reverse of the estimated shifts to the given projections to compensate for the shifts. Then, for each projection, we perform a local grid search around the current angle estimate to identify the angle at which the projection of the current image estimate best aligns—measured by the mean squared error—with the shifted input projection. 

A pseudo-code for this alternating minimization scheme is presented in Alg.~\ref{algo}. This method requires a specific form of angle initialization, explained in the next sub-section.

\begin{algorithm}[!t]
\caption{Three-Way Alternating Minimization}
\label{algo}
\begin{algorithmic}[1]
\State \textbf{Input:} Projections $\{\boldsymbol{y}_i\}_{i=1}^N$, initial angles $\{\hat{\theta}_i^{(0)}\}_{i=1}^N$ obtained from shift-aware Laplacian eigenmaps (cf Sec.~\ref{subsec:initangles}), initial shifts $\{\hat{\alpha}_i^{(0)}\}_{i=1}^N$ all set to 0
\State \textbf{Parameters:} Max iterations $T$, angle step $\delta$, number of angle trials $n$, tolerance $\varepsilon$
\State $\boldsymbol{\hat{f}}_{\text{prev}} \gets$ image recons. by FBP with initial angles and shifts
\For{$t = 1,\dots,T$}
    \State \textbf{Shift Estimation:}
    \For{$i = 1,\dots,N$}
        \State $\boldsymbol{\hat{y}}_i \gets \mathrm{Proj}(\boldsymbol{\hat{f}}^{(t-1)},\,\hat{\theta}_i^{(t-1)})$ 
        \State $\hat{\alpha}_i^{(t)} \gets \arg\max_{\alpha}\, \boldsymbol{y}_i \circledcirc \;\text{shift}_\alpha(\boldsymbol{\hat{y}}_i)$
    \EndFor

    \State \textbf{Image Reconstruction:}
    \State $\tilde{\boldsymbol{y}}_i \gets \text{shift}_{-\hat{\alpha}_i^{(t)}}(\boldsymbol{y}_i) \quad \forall i$
    \State $\boldsymbol{\hat{f}}^{(t)} \gets \mathrm{FBP}\big(\{\tilde{\boldsymbol{y}}_i\}_{i=1}^N,\;\{\hat{\theta}_i^{(t-1)}\}_{i=1}^N\big)$

    \State \textbf{Angle Update:}
    \For{$i = 1,\dots,N$}
        \For{$j = 1,\dots,n$}
            \State $\theta_{i,j} = \hat{\theta}_i^{(t-1)} + \delta\cdot\big(j - \lfloor n/2 \rfloor\big)$
            \State $\boldsymbol{\hat{y}}_{i,j} \gets \mathrm{Proj}(\boldsymbol{\hat{f}}^{(t)},\,\theta_{i,j})$
            \State $\mathcal{L}_{i,j} \gets \| \boldsymbol{y}_i - \text{shift}_{\hat{\alpha}_i^{(t)}}(\boldsymbol{\hat{y}}_{i,j}) \|_2^2$
        \EndFor
        \State $\hat{\theta}_i^{(t)} \gets \arg\min_j \mathcal{L}_{i,j}$
    \EndFor

    \State \textbf{Convergence Check:}
    \If{$\|\boldsymbol{\hat{f}}^{(t)} - \boldsymbol{\hat{f}}_{\text{prev}}\|_2 < \varepsilon$}
        \State \textbf{break}
    \EndIf
    \State $\boldsymbol{\hat{f}}_{\text{prev}} \gets \boldsymbol{\hat{f}}^{(t)}$
\EndFor
\State \textbf{Output:} $\boldsymbol{\hat{f}}^{(t)}$, $\{\hat{\alpha}_i^{(t)}\}_{i=1}^N$, $\{\hat{\theta}_i^{(t)}\}_{i=1}^N$
\end{algorithmic}
\end{algorithm}

\subsection{Initialization of angles} 
\label{subsec:initangles}
The performance of the alternating minimization scheme described earlier is sensitive to the initial angle estimates. A poor initialization can lead to convergence to an undesirable local minimum. To mitigate this, we propose a robust initialization method based on the graph Laplacian approach from \cite{coifman2008graph}, but \emph{modified} to handle the unknown spatial shifts in the projections. In the standard method, the distribution of the angles is assumed to be known (usually considered to be $\text{Uniform}(0,2\pi)$) even though the angles themselves are unknown. The projection angles are estimated by (\textit{i}) first circularly sorting the projections using dimensionality reduction, and (\textit{ii}) then assigning the angles using expected value of the order statistics of the angle distribution. For step (\textit{i}), an $N \times N$ similarity matrix $\boldsymbol{W}$ is generated where $W_{ij}$ is a similarity measure between $\boldsymbol{y}_i$ and $\boldsymbol{y}_j$, usually of the form $W_{ij} = e^{-\kappa\|\boldsymbol{y_i}-\boldsymbol{y_j}\|^2_2}$ for some parameter $\kappa > 0$. Using this similarity matrix $\boldsymbol{W}$, the Laplacian matrix $\boldsymbol{L}$ is created. The projections $\{\boldsymbol{y}_i\}_{i=1}^N$ are then projected onto the eigenvectors of $\boldsymbol{L}$ corresponding to the two smallest nonzero eigenvalues. Thus, each projection $\boldsymbol{y}_i$ is mapped to $\boldsymbol{\psi}_i \in \mathbb{R}^2$. The projections are sorted based on the angles formed by $\tan^{-1}(\psi_{i,1}/\psi_{i,2})$. Note that these angles do not represent the actual projection angles and are only used for the purposes of circular sorting. Once the angles are sorted, angle assignment is done using order statistics. For $\text{Uniform}(0,2\pi)$, the angle for the $i$th projection ($i \in \{1,...,N\}$) in the ordering is set to be $\theta_i := 2\pi (i-1)/N$. 

However, in the presence of shifts in projections, the similarity values in $\boldsymbol{W}$ require modification to accurately reflect similarity. Thus, a crucial step in our modification is the construction of a shift-aware similarity matrix $\boldsymbol{W}$. For this, define $W_{ij}$ as the similarity metric applied between $\boldsymbol{y}_i$ and a shifted version of $\boldsymbol{y}_j$, where the shift in $\boldsymbol{y}_j$ is chosen such that the dot product between $\boldsymbol{y}_i$ and shifted $\boldsymbol{y}_j$ is maximized. That is, 
$W_{ij} := \text{sim}(\boldsymbol{y}_i, \text{shift}_{k_0}(\boldsymbol{y}_j))$,
where $k_0 = \argmax_{k} \quad \boldsymbol{y}_i  \circledcirc\text{shift}_k(\boldsymbol{y}_j)$. Here $\text{sim}(\cdot, \cdot)$ represents the similarity measure and $\text{shift}_k(\boldsymbol{y}_j)$ represents the shifted version of $\boldsymbol{y}_j$, shifted by $k$.
Once this modified $\boldsymbol{W}$ is obtained, we continue with the standard procedure by obtaining the Laplacian matrix as outlined in \cite{coifman2008graph}.

\section{Results}
\label{sec:results}
We evaluated our method using two different images of size $256 \times 256$. The images are 2D slices of 3D maps of the \textit{Cutibacterium acnes} 70S ribosome (Image 1) \cite{emd26959} and the \textit{Leishmania Major} 80S ribosome (Image 2) \cite{emdb50739}. For each ground truth image, we generated 3000 projections at angles sampled from $\text{Uniform}(0, 2\pi)$.
To simulate realistic acquisition errors, two types of distortions were introduced. First, prior to generating the $i$-th projection, the ground truth image was shifted by a random offset $(s_0^i, t_0^i)$, where both $s_0^i$ and $t_0^i$ were integers sampled uniformly randomly from the set $\{-M, \dots, M\}$ where we varied $M$ from $5$ to $15$. Second, noise was added to each projection, drawn from a distribution $\mathcal{N}(0, \sigma^2)$, where the standard deviation $\sigma$ was set to a fraction $\gamma$ of the mean absolute value of the clean projections. We also varied the value of $\gamma$ from $0.05$ to $0.07$ in our experiments.

\begin{figure*}[h]
\centering

\begin{tabular}{c}

    \begin{tabular}{@{}c@{\hspace{2pt}} *{4}{c@{\hspace{2pt}}} } %
    & Original Image & Blind & Ours & Oracle \\

    \begin{turn}{90}\hspace{40pt}Image 1\end{turn} &
    \includegraphics[width=0.23\textwidth, height=0.20\textwidth, keepaspectratio]{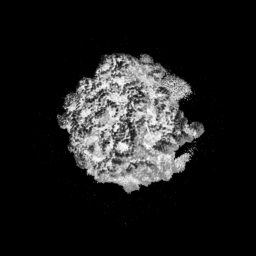} &
    \includegraphics[width=0.23\textwidth, height=0.20\textwidth, keepaspectratio]{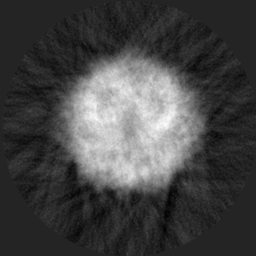} &
    \includegraphics[width=0.23\textwidth, height=0.20\textwidth, keepaspectratio]{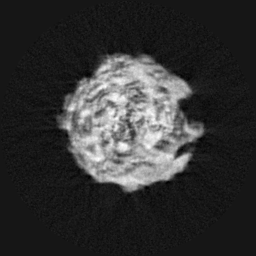} &
    \includegraphics[width=0.23\textwidth, height=0.20\textwidth, keepaspectratio]{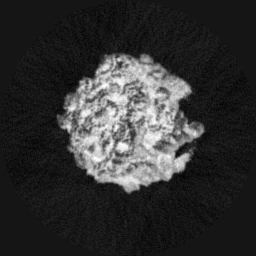} \\

    \begin{turn}{90}\hspace{40pt}Image 2\end{turn} &
    \includegraphics[width=0.23\textwidth, height=0.20\textwidth, keepaspectratio]{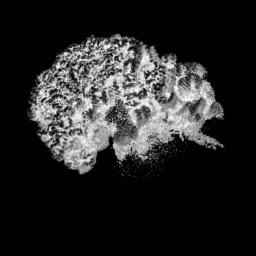} &
    \includegraphics[width=0.23\textwidth, height=0.20\textwidth, keepaspectratio]{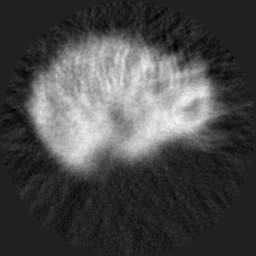} &
    \includegraphics[width=0.23\textwidth, height=0.20\textwidth, keepaspectratio]{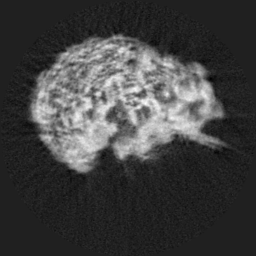} &
    \includegraphics[width=0.23\textwidth, height=0.20\textwidth, keepaspectratio]{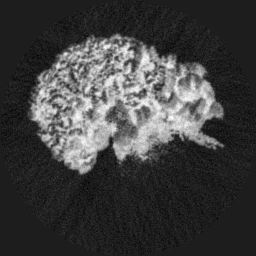} \\

    \end{tabular}
\end{tabular}
\caption{Visual comparison between the ground truth and reconstructions from 3000 noisy projections, obtained from our algorithm, Blind reconstruction as in \cite{coifman2008graph}, and Oracle for $\gamma = 0.06$ and $M = 10$. Our technique is able to reconstruct the image much better than the Blind reconstruction.}
\label{fig:visual_comparison}
\end{figure*}

\begin{figure}
    \centering
    \includegraphics[width=0.9\linewidth]{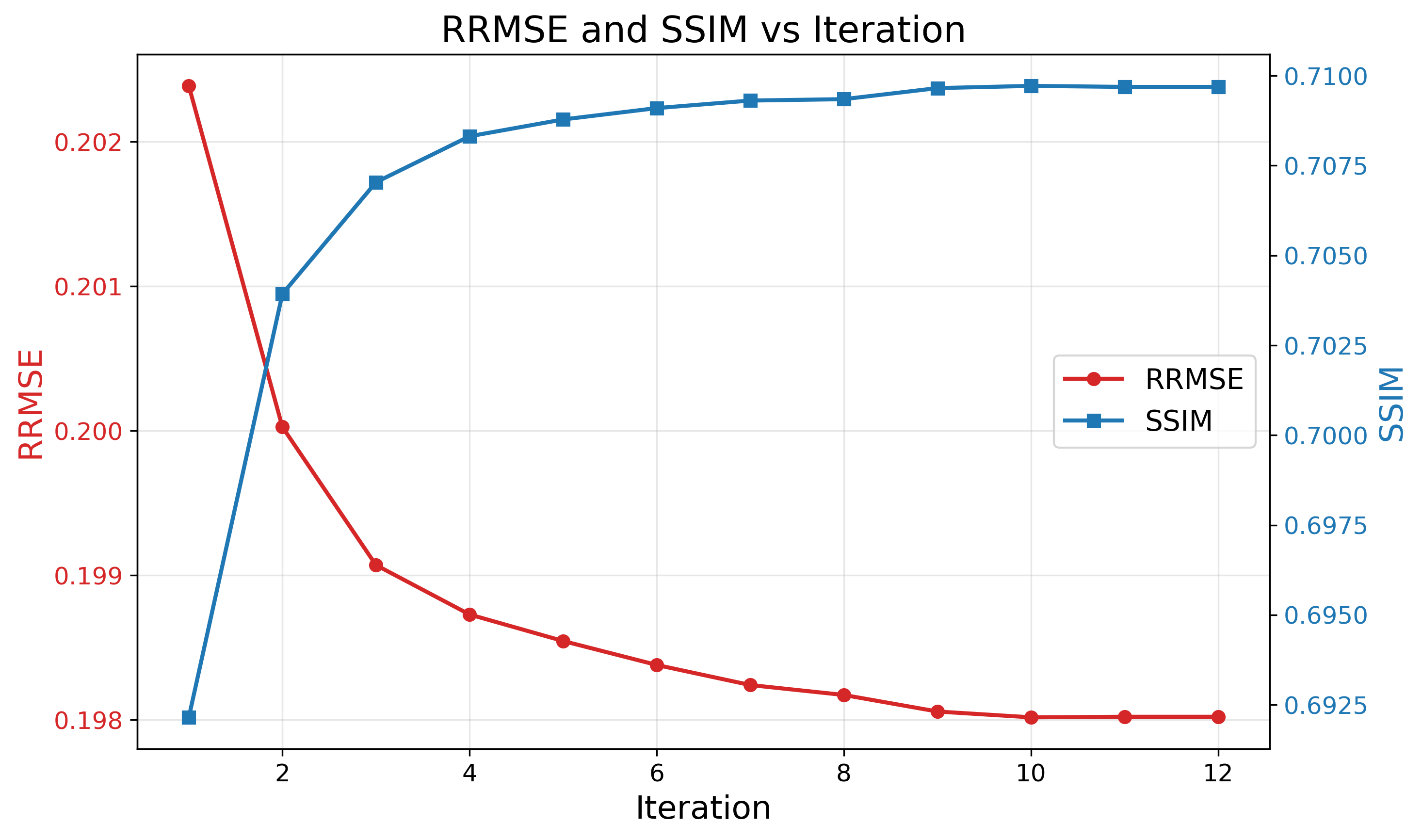}
    \caption{Variation of RRMSE and SSIM across iterations for Image 1 for $M = 10$ and $\gamma = 0.06$.}
    \label{fig:convergence}
\end{figure}

The performance of our algorithm was compared against two baselines defined as follows: (1) \textbf{Blind:} Angles are initialized using the graph Laplacian-based method from \cite{coifman2008graph}. Then, FBP is used to reconstruct the image, assuming no shifts. (2) \textbf{Oracle:} Projections are shifted back using the ground truth shifts, and ground truth angles are used to reconstruct the image using FBP. This gives an upper bound on the reconstruction quality that can be achieved using FBP. To evaluate our reconstruction, we align our reconstructed image $\boldsymbol{\hat{f}}$ with the ground truth image $\boldsymbol{f}$ and compute the relative root mean-squared-error (RRMSE) $\|\boldsymbol{\hat{f}}-\boldsymbol{f}\|_2/\|\boldsymbol{f}\|_2$, the correlation coefficient (CC) and the structural similarity index (SSIM) \cite{wang2009mean} between them. The quantitative results, measured using these metrics, of our proposed method for various distortion levels is presented in Table~\ref{tab:results}.
The qualitative impact of our method for $M = 10$ and $\gamma = 0.06$ is shown in Figure \ref{fig:visual_comparison}. It is clear that the final aligned result of our algorithm is significantly sharper and more detailed than the baseline reconstruction. For the same values of $M$ and $\gamma$, the plot in Figure \ref{fig:convergence} shows the reduction in RRMSE and increment SSIM in across iterations of Alg.~\ref{algo}.



\begin{table*}[h]
\centering
\caption{Performance comparison of Blind, proposed, and Oracle algorithms under different settings of $\gamma$ and $M$ for Image 1.}
\label{tab:results}
\begin{tabular}{|c|c|c|c|c|c|c|c|c|c|c|}
\hline
\multirow{2}{*}{$\gamma$} & \multirow{2}{*}{$M$} & \multicolumn{3}{|c|}{RRMSE} & \multicolumn{3}{|c|}{SSIM} & \multicolumn{3}{|c|}{CC} \\
\cline{3-11}
& & \textcolor{blind_color}{Blind} & \textcolor{ours_color}{Ours} & \textcolor{oracle_color}{Oracle} & \textcolor{blind_color}{Blind} & \textcolor{ours_color}{Ours} & \textcolor{oracle_color}{Oracle} & \textcolor{blind_color}{Blind} & \textcolor{ours_color}{Ours} & \textcolor{oracle_color}{Oracle} \\
\hline
0.05 & 5 & \textcolor{blind_color}{0.416} & \textcolor{ours_color}{0.176} & \textcolor{oracle_color}{0.120} & \textcolor{blind_color}{0.386} & \textcolor{ours_color}{0.747} & \textcolor{oracle_color}{0.677} & \textcolor{blind_color}{0.887} & \textcolor{ours_color}{0.981} & \textcolor{oracle_color}{0.991} \\
\hline
0.06 & 10 & \textcolor{blind_color}{0.410} & \textcolor{ours_color}{0.198} & \textcolor{oracle_color}{0.125} & \textcolor{blind_color}{0.340} & \textcolor{ours_color}{0.710} & \textcolor{oracle_color}{0.651} & \textcolor{blind_color}{0.890} & \textcolor{ours_color}{0.976} & \textcolor{oracle_color}{0.990} \\
\hline
0.07 & 15 & \textcolor{blind_color}{0.403} & \textcolor{ours_color}{0.188} & \textcolor{oracle_color}{0.131} & \textcolor{blind_color}{0.390} & \textcolor{ours_color}{0.709} & \textcolor{oracle_color}{0.614} & \textcolor{blind_color}{0.895} & \textcolor{ours_color}{0.978} & \textcolor{oracle_color}{0.989} \\
\hline
\end{tabular}
\end{table*}

\section{Conclusion}
\label{sec:conclusion}
In this paper, we have presented a method for 2D tomographic reconstruction when faced with the dual challenges of unknown viewing angles and unknown spatial shifts in the projection data. Our approach integrates a shift-aware initialization technique, which modifies an existing graph Laplacian-based algorithm, with a three-way alternating minimization scheme. This allows for the joint and iterative refinement of the image, its projection angles, and the corresponding shifts. The experimental results, conducted on noisy projections of ribosome images, demonstrate the effectiveness of our proposed algorithm. Compared to a baseline method that does not account for spatial shifts, our technique achieves significantly sharper reconstructions, as validated by quantitative metrics such as RRMSE, SSIM, and CC. Our method can be easily integrated with a technique that estimates the unknown distribution of the angle directly from the projections \cite{shah2025unknown,adversarial2022}.

\newpage
\bibliographystyle{IEEEbib}
\bibliography{strings,refs}

\section{Comparison with Moment-Based Angle Recovery}

In response to reviewer suggestions, we additionally compared our method with the recent moment-based approach for joint angle and shift recovery proposed by Hjouj et al.~\cite{Hjouj2025-mr}. This method estimates projection angles by analytically eliminating unknown shifts using first-order moments, followed by candidate angle recovery from second-order moments and disambiguation via third-order moments.

\subsection{Methodological differences}
We first note a fundamental difference between the two approaches. The method of \cite{Hjouj2025-mr} estimates each projection angle independently using low-order global moments, and therefore does not exploit redundancy across projections. In contrast, our approach leverages consistency across all projections which provides substantial robustness to noise and discretization.

Moreover, the moment-based method requires access to projections at four known reference angles $\{0, \pi/4, \pi/2, 3\pi/4\}$ to estimate the image moments used for angle recovery. These reference projections are assumed to be accurately centered and noiseless. Such assumptions are difficult to satisfy in practice, where projection angles are fully unknown and measurements are heavily corrupted by noise and spatial shifts.

\subsection{Empirical comparison}
We implemented the method of \cite{Hjouj2025-mr} on Image 1 under the same experimental setting as our approach, using the same image-domain shift model, discretized Radon transform, noise model, and reconstruction operators.

Even in the absence of additive noise ($\gamma = 0$), the moment-based method produces reconstructions with an RRMSE of $0.340$, SSIM of $0.281$, and CC of $0.938$ for $M=5$. These values are worse than those obtained by our method at $\gamma=0.05$ (see Table~\ref{tab:results}).

\subsection{Angle estimation behavior}
\begin{figure}[t]
    \centering
    \includegraphics[width=0.9\linewidth]{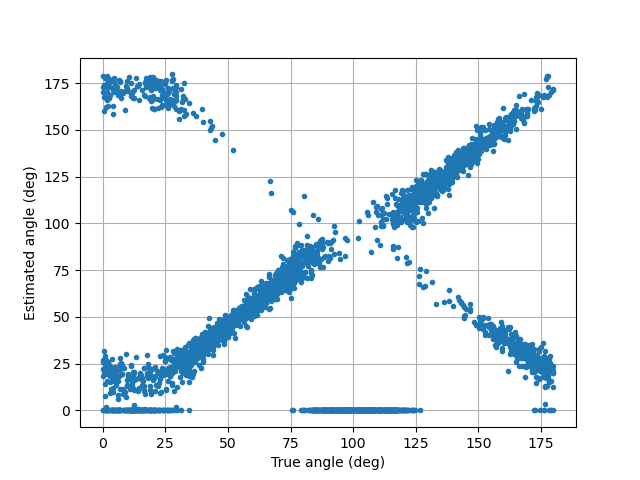}
    \caption{Estimated angles versus ground-truth angles for the moment-based method of~\cite{Hjouj2025-mr} at $\gamma = 0.01$.
Second-order moments recover projection angles only up to a reflection ambiguity, producing an X-shaped pattern. Third-order moments, which are required to resolve this ambiguity, fail to consistently select the correct branch.}
    \label{fig:hjouj_angles}
\end{figure}

Figure~\ref{fig:hjouj_angles} shows the estimated angles versus ground-truth angles for the moment-based method at $\gamma=0.01$. The X-shaped pattern indicates that second-order moments recover angles only up to a reflection ambiguity, while the third-order moment used for disambiguation becomes unstable, and is unable to consistently correctly resolve the ambiguity. As noise increases further ($\gamma=0.05$), we observe that even first- and second-order moments become unreliable, leading to near-random angle estimates.

These observations highlight the sensitivity of moment-based angle recovery to discretization, interpolation, and noise, as well as its reliance on special reference projections. In contrast, our method exploits redundancy across projections and remains stable under these effects.

\end{document}